\documentclass[aps,pra,twocolumn,showpacs,superscriptaddress]{revtex4-1}

\usepackage{graphicx}
\usepackage[utf8]{inputenc}
\usepackage[english]{babel}
\usepackage[colorlinks,citecolor=blue,urlcolor=blue]{hyperref}
\usepackage{amsmath}
\usepackage{amssymb}
\usepackage{epsfig}

\newcommand{\Tr}{\mathrm{Tr}}
\newcommand{\diag}{\mathrm{diag}}
\newcommand{\eq}[1]{(\ref{#1})}

\begin{document}

  \title{Transformations of symmetric multipartite Gaussian states by Gaussian LOCC}

  \author{Ond\v{r}ej \v{C}ernot\'ik}
  \email{Ondrej.Cernotik@itp.uni-hannover.de}
  \affiliation{Department of Optics, Palack\'y University, 17. listopadu 12, 77146 Olomouc, Czech Republic}
  \affiliation{Institute for Theoretical Physics, Leibniz University Hannover, Appelstra\ss{}e 2, 30167 Hannover, Germany}
  \affiliation{Max Planck Institute for Gravitational Physics (Albert Einstein Institute),
     Callinstra\ss{}e 38, 30167 Hannover, Germany}

  \author{Jarom\'ir Fiur\'a\v{s}ek}
  \affiliation{Department of Optics, Palack\'y University, 17. listopadu 12, 77146 Olomouc, Czech Republic}

  \date{\today}

  \begin{abstract}
     Multipartite quantum correlations, in spite of years of intensive research, still leave many questions unanswered.
     While bipartite entanglement is relatively well understood for Gaussian states,
     the complexity of mere qualitative characterization grows rapidly with increasing number of parties.
     Here, we present two schemes for transformations of multipartite permutation invariant Gaussian states
     by Gaussian local operations and classical communication.
     To this end, we use a scheme for possible experimental realization,
     making use of the fact, that in this picture, the whole $N$-partite state can be described
     by specifying the states of two separable modes.
     Numerically, we study entanglement transformations of tripartite states.
     Finally, we look at the effect our protocols have on fidelity of assisted quantum teleportation and find
     that while adding correlated noise does not affect the fidelity at all,
     there is strong evidence that partial non-demolition measurement leads to a drop in teleportation fidelity.
  \end{abstract}

  \pacs{03.67.Bg, 42.50.Ex, 03.67.Ac}

  \maketitle

  \section{Introduction}

  Gaussian states and operations, though a mere subset of continuous-variable systems,
  represent an important resource in quantum information processing and,
  particularly, in quantum optics~\cite{Braunstein05,Weedbrook12}.
  Their importance lies in the fact that they can be easily described using only the displacement and covariance matrix,
  the first and second statistical moments of the quadrature operators,
  making use of the formalism based on symplectic analysis;
  on the other hand, their experimental importance stems from the fact that Gaussian states can be generated and manipulated 
  using coherent laser light, passive linear optical elements, optical parametric amplifiers
  and highly efficient homodyne detection.
  All these tools enable us to generate Gaussian entanglement deterministically.

  The significance of Gaussian states is further emphasized when studying entanglement.
  Compared to a general system of two $N$-level systems,
  where the question of bound entanglement is still an important topic of research,
  it has been shown that with two-mode bipartite Gaussian states
  separability is equivalent to positive partial transpose~\cite{Duan00,Simon00},
  and later it has been proved that this holds for any $1\times N$-mode bipartite states~\cite{Werner01}.

  Multipartite Gaussian entanglement~\cite{Aoki03}, nevertheless, still represents a considerable challenge
  in our understanding of quantum correlations.
  Even restricting to scenarios, where each party is in possession of only a single mode,
  the complexity of the system grows rapidly with increasing number of parties.
  While two-mode bipartite Gaussian states are either entangled or separable
  (depending on the positivity of the partial transpose, therefore easily determined),
  with three-mode tripartite entangled Gaussian states, there are five entanglement classes,
  depending on the separability of all three possible bipartitions of the state~\cite{Giedke01},
  and, to the best of our knowledge, no such simple classification exists for Gaussian states of four parties.
  Still, sufficient criteria for multipartite continuous-variable entanglement have been derived~\cite{Loock03,Reid13}
  and additional methods for characterizing multipartite entanglement were proposed,
  e.g., using localization to two-mode entanglement~\cite{Serafini05,Fiurasek07},
  connection between multipartite entanglement and teleportation fidelity~\cite{Adesso05},
  or using multipartite entanglement witnesses~\cite{Sperling13}.
  A need for better understanding of multipartite Gaussian entanglement is, moreover, motivated by recent advances
  in experimental generation of Gaussian cluster states~\cite{Yukawa08,Yokoyama13,Roslund13},
  and by recent experimental demonstration of entanglement distribution using separable states~\cite{Peuntinger13,Vollmer13}.

  In this paper, we analyze protocols for transformations of multipartite Gaussian states by local Gaussian operations and
  classical communication.
  We are mainly interested in a qualitative characterization of entanglement;
  therefore, our numerical results concern only tripartite Gaussian states that still can be unambiguously classified,
  adopting the terminology of Ref.~\cite{Giedke01}.
  In addition, this approach is similar to that of Giedke and Kraus~\cite{Giedke13}
  who were, nevertheless, interested in a more general equivalence of $N$-mode entangled Gaussian states
  while we propose specific protocols to achieve this task.
  Moreover, the focus of Ref.~\cite{Giedke13} lies in Gaussian local unitaries,
  whereas our protocols use a wider class of Gaussian local operations and classical communication.

  Secondly, our motivation is also to generalize protocols for full symmetrization of bipartite Gaussian
  states~\cite{Fiurasek12}.
  Our generalization is twofold---not only do we consider a higher number of modes
  but we also relax the condition of full symmetry.
  By fully symmetric, we mean states that are not only invariant with respect to the exchange of the two modes,
  but also have equal amplitude and phase variances and exhibit equally strong correlations in both quadratures.
  Our generalization works again with permutation symmetric states,
  i.e., states that are not changed by exchanging any two modes,
  however, we do not require equal variances nor correlations.

  The rest of the paper is organized as follows:
  We review the description of Gaussian states and operations in Sec.~\ref{sec.math}.
  Here, we also present a scheme of equivalent state preparation
  that enables us to describe the $N$-partite states using only two separable modes.
  Individual strategies used for transformations of symmetric Gaussian states are introduced in the following sections.
  Specifically, protocol based on correlated noise addition is studied in Sec.~\ref{sec.noise},
  and the use of partial non-demolition measurement is investigated in Sec.~\ref{sec.qnd}.
  In Sec.~\ref{sec.teleportation}, we study assisted quantum teleportation with permutation invariant Gaussian states
  and investigate the effect of the aforementioned protocols on the teleportation fidelity.
  Finally, we conclude in Sec.~\ref{sec.conclusions}.

  \section{Mathematical prerequisites}\label{sec.math}

  Starting from creation $\hat{a}^\dagger_j$ and annihilation $\hat{a}_j$ operators,
  we can introduce the amplitude and phase quadratures as
  $\hat{x}_j = \hat{a}_j+\hat{a}_j^\dagger$, $\hat{p}_j = i(\hat{a}_j^\dagger-\hat{a}_j)$.
  Collecting the quadrature operators of $N$ modes into a vector $\hat{r} = (\hat{x}_1,\hat{p}_1,\ldots,\hat{x}_N,\hat{p}_N)^T$,
  we can write the commutation relations using the symplectic form~\cite{Weedbrook12} $\Omega = \bigoplus_{j=1}^N \omega$,
  \begin{equation}
    \omega = \left(\begin{array}{cc} 0&1\\-1&0 \end{array}\right),
  \end{equation}
  as $[\hat{r}_j,\hat{r}_k] = 2i \Omega_{jk}$.
  Gaussian states, i.e., states with a Gaussian phase-space representation (e.g., the Wigner function),
  are then described using first and second statistical moments of the quadrature operators,
  the mean value $\bar{r} = \Tr(\hat{r}\hat{\rho})$,
  and the covariance matrix with elements $\gamma_{jk} = \langle\{\Delta\hat{r}_j,\Delta\hat{r}_k\}\rangle/2$,
  where $\Delta\hat{r}_j = \hat{r}_j-\langle\hat{r}_j\rangle$ and $\{,\}$ denotes the anticommutator.
  Note that the mean value can be changed deterministically using local displacements
  and does not affect the entanglement of the state.
  We will, therefore, use only the covariance matrix to describe Gaussian states;
  we will often use the covariance matrix as a full state description instead of the density matrix,
  speaking of state $\gamma$ when referring to state $\hat{\rho}$.
  Gaussian unitary operations, i.e., unitaries that map Gaussian states to Gaussian states,
  can be described by their action on the covariance matrix,~\cite{Weedbrook12,Braunstein05}
  \begin{equation}
    \gamma \to S\gamma S^T,
  \end{equation}
  where $S$ is a symplectic matrix;
  general Gaussian completely positive maps can, nevertheless, also be described using the covariance matrix formalism~\cite{Fiurasek02b,Giedke02}.

  In the following, we will consider permutationally invariant Gaussian states with the covariance matrix
  \begin{equation}
    \gamma = \left(\begin{array}{cccc} \nu&\sigma&\ldots&\sigma \\ \sigma^T&\nu&\ldots&\sigma \\
      \vdots&&\ddots&\vdots \\ \sigma^T&\ldots&\sigma^T&\nu \end{array}\right).
  \end{equation}
  Here, $\nu$ and $\sigma$ are 2$\times$2 matrices describing the mode covariance and the inter-modal correlations, 
  respectively.
  We assume canonical form, with diagonal $\nu$ and $\sigma$;
  as a consequence, these matrices can be parametrized either by the specific variances and correlations,
  $\nu = \diag(m,n)$, $\sigma = \diag(c,-d)$,
  or using the ratio of the diagonal terms, $\nu = \diag(m,k_1m)$, $\sigma = \diag(c,-k_2c)$.

\begin{figure}
    \centering
    \includegraphics[width=\linewidth]{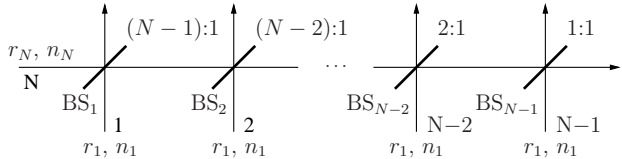}
    \caption{\label{fig.scheme}
      An effective scheme for preparation of the studied states.
      The input mode $N$ is distributed equally among all output ports, resulting in a permutationally invariant output state.
      The main reason for introducing the scheme is, however, to simplify the analysis of the investigated protocols.
    }
  \end{figure}

  In Fig. \ref{fig.scheme}, we show how to prepare such states in an experimental realization.
  Such a scheme was originally proposed by van Loock and Braunstein~\cite{Loock00} for three modes
  and later used in experiments demonstrating generation of multipartite Gaussian entanglement~\cite{Aoki03}
  and assisted quantum teleportation~\cite{Yonezawa04}.
  The $N$ modes, $N-1$ of which are in an identical thermal squeezed state with noise $n_1$ and squeezing $r_1$
  while the remaining mode has thermal noise $n_N$ and squeezing $r_N$,
  are superimposed on an array of beam splitters with transmittance-reflectance ratios $(N-1):1$, \ldots, $1:1$.
  The effect of this setup is to distribute the $N$-th mode (i.e., the only different one) equally among all output modes.
  Denoting the variances of the input quadratures 
  $V_x = n_1e^{2r_1}$, $V_p = n_1e^{-2r_1}$, $W_x = n_Ne^{2r_N}$, $W_p = n_Ne^{-2r_N}$,
  and assuming the beam splitters imprint a phase shift of $\pi$ on modes 1,\ldots,$N-1$ upon reflection,
  it is a straightforward task to show 
  that the following relations between input variances and parameters of the output covariance matrix hold
  \begin{eqnarray}
    m &=& [(N-1)V_x+W_x]/N,\label{eq.m} \\
    n &=& [(N-1)V_p+W_p]/N, \\
    c &=& (W_x-V_x)/N, \label{eq.c}\\
    d &=& (V_p-W_p)/N, \label{eq.d}
  \end{eqnarray}
  or, inversely,
  \begin{eqnarray}
    V_x &=& m-c, \\
    V_p &=& n+d, \\
    W_x &=& m+(N-1)c, \\
    W_p &=& n-(N-1)d.
  \end{eqnarray}

  With the use of the presented experimental scheme, 
  we can significantly simplify the investigation of the protocols used for transformations of the states.
  Denoting the covariance matrix of the input state of the $N$-port beam splitter as $\gamma_\mathrm{in}$ 
  and using $B$ to describe the $N$-port beam splitter transformation,
  we have the relation $\gamma = B\gamma_\mathrm{in}B^T$.
  If we consider a symplectic transformation $S$ on the state $\gamma$ and use the fact that $BB^T = B^TB = I$, 
  where $I$ is the identity, we can write
  \begin{equation}
    S\gamma S^T = SB\gamma_\mathrm{in}B^TS^T = BB^TSB\gamma_\mathrm{in}B^TS^TBB^T.
  \end{equation}
  Instead of performing the operation $S$ on the state $\gamma$ (which would be, in an experiment, 
  prepared by mixing the modes of the state $\gamma_\mathrm{in}$ on the $N$-port beam splitter $B$), 
  we can thus perform the operation $B^TSB$ on the state $\gamma_\mathrm{in}$ 
  and then mix its modes on the $N$-port beam splitter $B$.
  As a result, this description enables us to consider the effect of the transformation $B^TSB$ on the separable state $\gamma_\mathrm{in}$
  (which will, in most cases, result in local transformations on each mode) 
  instead of calculating the overall transformation $S$ on the whole (possibly entangled) state $\gamma$.
  Moreover, if the operation factorizes into a product of identical local single-mode operations
  (as is often the case since we want to preserve the permutation invariance),
  the operation is unaffected by the $N$-port beam splitter 
  and it does not matter if we apply the transformation on the input or output modes
  as can be proved by writing the symplectic operation in block form
  \footnote{Operation that is local and the same on all modes is then block diagonal, $S = \diag(S_1,S_1,\ldots,S_1)$.
    In the block form, it is then straightforward to show that $B^TSB = S$ for this type of transformation.}.
  
  \begin{figure*}
    \centering
    \includegraphics[width=\linewidth]{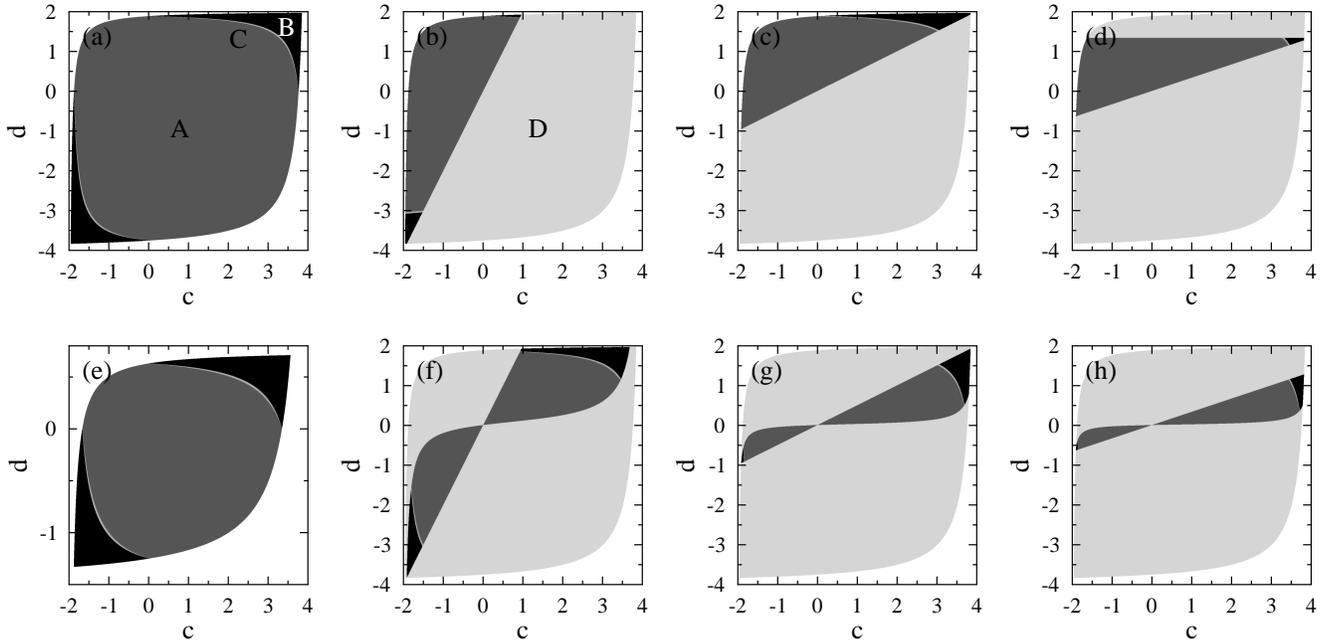}
    \caption{ \label{fig.classes}
       Left column: Entanglement classes for states with $m = n = 4$ (a) and $m = 4$, $n = 1.5$ (e).
       In the remaining panels, transformations of entanglement classes 
       with the noise addition protocol [top row, (b)-(d)] and with the QND interaction strategy [bottom, (f)-(h)] are shown.
       Area A (dark gray) represents fully separable states of class V,
       area B (black) shows fully entangled states, i.e., states belonging to class I
       while in area C (gray), bound entangled states, or states of class IV are shown.
       Area D (large areas of light grey) shows states that cannot be brought to a state with required
       values of $k_{1,2}$ using given protocol and unphysical states are shown in white.
       Note that although the set of bound entangled states is very small compared to both classes I, V, 
       it has a non-zero measure.
       The required output state ratios $k'_{1,2}$ in panels (b)-(d), (f)-(h) are
       $k'_1 = 1$, $k'_2 = 2$ for panels (b), (f); $k'_1 = 2$, $k'_2 = 1$ (c), (g);
       $k'_1 = 3$, $k'_2 = 1$ (d), (h); moreover, numerical calculations suggest that only the ratio $k'_2/k'_1$ is relevant,
       i.e., parameters $k'_1 = 2$, $k'_2 = 4$ would reproduce the results of panels (b), (f).
       Finally, the behavior with different values of input variances $m$, $n$ is qualitatively the same.
    }
  \end{figure*}

  To qualitatively characterize entanglement of the input and output tripartite states, we follow the approach of Ref.~\cite{Giedke01}.
  These states can belong to five entanglement classes, three of which are relevant for permutationally invariant states.
  The states can be fully entangled, i.e., inseparable with respect to any bipartition of the state, fully separable,
  or bound entangled, i.e., separable with respect to any bipartition, yet not separable to product state of three subsystems.
  Adopting the terminology of Ref.~\cite{Giedke01}, the states are said to belong to entanglement class I, V and IV, respectively.
  As an example, in Fig.~\ref{fig.classes}, we show entanglement classes for states with $m = n = 4$ (a)
  and $m = 4$, $n = 1.5$ (e).
  While it is not clearly visible in the figure, bound entangled states form a boundary between fully separable and fully entangled states of a finite width.
  Thus, it is in principle possible to create these states in a laboratory, given the experimental error is sufficiently small.

  \section{Correlated noise addition}\label{sec.noise}

  The first approach for transformations of symmetric Gaussian states we study is based on adding correlated noise to each mode of the $N$-partite state.
  This can, in practice, be accomplished performing random local displacements $x_j\to x_j+x_n$,
  where $x_n$ is a Gaussian random variable with zero mean and variance $V_N/2$.
  The effect of the noise addition, in terms of covariance matrices, is given by
  \begin{equation}
    \gamma\to\gamma+\gamma_n = 
    \gamma+\left(\begin{array}{ccc} \alpha&\alpha&\ldots \\ \alpha&\alpha&\ldots \\ \vdots&&\ddots \end{array}\right),
  \end{equation}
  where $\gamma$ is the covariance matrix of the input state and $\gamma_n$ describes the added noise;
  it is composed of 2$\times$2 blocks $\alpha = \diag(V_N,0)$.
  We can see that the added noise increases correlations between amplitude quadratures of individual modes;
  if the amplitude quadratures are initially correlated, i.e., $c>0$ in off-diagonal blocks $\sigma = \diag(c,-d)$ in the original covariance matrix,
  these correlations are increased.
  On the other hand, the noise also increases variance of the amplitude quadrature of each mode.
  As we will see in Sec.~\ref{sec.teleportation}, these two effects exactly cancel each other when using the state for quantum teleportation.
  Thus, this procedure does not increase quantum correlations between modes, as one might naively expect from the increase of
  correlations~$c$.

  \begin{figure}
    \centering
    \includegraphics[width=0.75\linewidth]{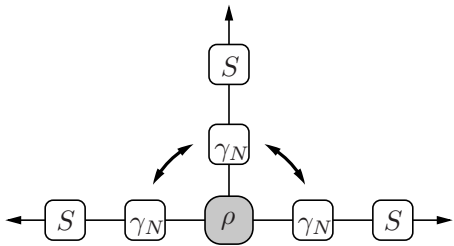}
    \caption{\label{fig.noise}
      Protocol for transformations of permutation symmetric Gaussian states by adding correlated noise.
      The parties first add correlated Gaussian noise with zero mean and variance $V_N/2$ to the amplitude quadrature of their mode
      and subsequently apply local squeezing to their mode.
    }
  \end{figure}


  After adding correlated noise, all parties perform additional squeezing on their mode;
  the whole protocol is sketched in Fig.~\ref{fig.noise}.
  Using the effective scheme in Fig.~\ref{fig.scheme}, the protocol corresponds to adding Gaussian noise with variance $NV_N$ to mode N
  and performing squeezing on each mode.
  The former can be seen by propagating the noise covariance matrix $\gamma_N$ through the $N$-port beam splitter $B$ for three modes,
  the latter is a consequence of the fact that the squeezing is the same on each mode and is therefore, as discussed in Sec.~\ref{sec.math},
  invariant to the beam splitter transformation.
  Taking this into account, the transformation can be described in the effective scheme by
  \begin{eqnarray}
    V_x &\to& \frac{1}{a}V_x,\label{eq.noise_vx} \\
    V_p &\to& aV_p,\\
    W_x &\to& \frac{1}{a}(W_x+NV_N),\label{eq.noise_wx}\\
    W_p &\to& aW_p,\label{eq.noise_wp}
  \end{eqnarray}
  with $a = e^{2r}$ giving the squeezing.
  Using Eqs.~\eq{eq.m}-\eq{eq.d} and requiring the ratios $k'_1 = n'/m'$, $k'_2 = d'/c'$ in the output state
  (throughout the paper, we parametrize output states by primed quantities and input states by unprimed ones),
  we obtain the solution
  \begin{eqnarray}
    a^2 &=& \frac{k'_1k'_2NV_x}{[k'_2(N-1)-k'_1]V_p+(k'_1+k'_2)W_p},\nonumber \\
       &=& k'_1k'_2\frac{m-c}{k'_2n-k'_1d}, \label{eq.noise_squeezing}\\
    V_N &=& \frac{V_x-W_x}{N}+\frac{V_p-W_p}{k'_2N}a^2 = \frac{k'_1md-k'_2nc}{k'_2n-k'_1d},\label{eq.noise_noise}
  \end{eqnarray}
  the physicality of the squeezing and noise is given by the conditions
  \begin{equation}\label{eq.noise_physicality}
     \frac{nc}{m}\frac{k'_2}{k'_1} < d < n\frac{k'_2}{k'_1}.
  \end{equation}

  The permutation invariant multipartite Gaussian states are described using four real parameters, quadrature variances and correlations,
  while the transformation protocol has two degrees of freedom, the noise variance and the squeezing.
  When generalizing protocols for full symmetrization to operations preserving the permutation symmetry,
  it is therefore natural to vary the variance and correlation ratios $k'_{1,2}$.
  Nonetheless, it is also possible to choose a different pair of parameters that ought to be changed by the protocol.
  Consequently, our approach is of a more general applicability---the equations describing the effective scheme transformation \eq{eq.noise_vx}-\eq{eq.noise_wp}
  together with the relations between the effective scheme and original state \eq{eq.m}-\eq{eq.d} and the required pair of parameters
  can be used to give the required noise variance and squeezing.
  This generalization is also possible with the protocol based on partial quantum non-demolition measurement that is introduced in the next section.

  We study the transformations of entanglement classes of tripartite states by this protocol in
  Fig.~\ref{fig.classes} [top row, panels (b)-(d)].
  The physicality of squeezing and noise variance Eq.~\eq{eq.noise_physicality} affects the set of states
  that can be transformed by this protocol in a straightforward way:
  The first inequality gives rise to the linear cutoff in each panel with the slope given by the ratio $k'_2/k'_1$.
  The second inequality leads to the horizontal cutoff that appears for small ratio $k'_2/k'_1$ in panel (d).

  Secondly, for $c, d<0$, there is a rapid change in entanglement classes, cf. Fig.~\ref{fig.classes} (a).
  For $c,d>0$, on the other hand, the entanglement classification is unaffected by the transformation
  up to a very small widening of the set of bound entangled states at the expense of fully entangled states.
  In other words, while the boundary of fully separable and bound entangled states stays fixed,
  the border between bound entangled and fully entangled states moves slightly into the region of fully entangled states.
  We conclude this section by noting that for different values of amplitude and phase variances,
  the behavior is analogous in all respects discussed here.

  \section{Partial QND measurement}\label{sec.qnd}

\begin{figure}
    \centering
    \includegraphics[width=\linewidth]{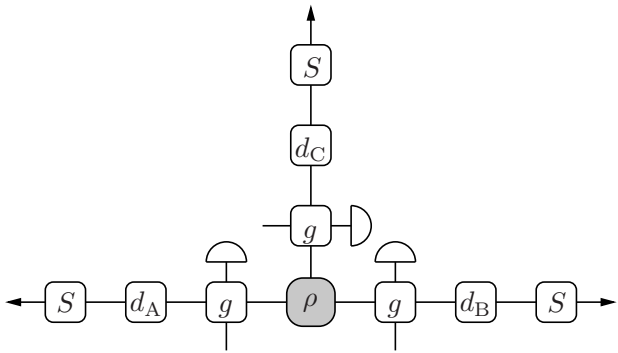}
    \caption{\label{fig.qnd}
      Scheme of protocol for symmetric Gaussian state transformation based on partial QND measurement.
      Each mode undergoes a non-demolition interaction of strength $g$ with ancillary mode that is subsequently measured.
      After the measurement results are made public, each party performs a displacement $d$ on their mode.
      In the end, each mode is subject to squeezing $S$.
    }
  \end{figure}

  The second strategy, which is schematically illustrated in Fig.~\ref{fig.qnd}, is based on a quantum non-demolition (QND) interaction with auxiliary modes,
  initially in the vacuum state.
  The ancillas are subsequently measured and, depending on the measurement results that are announced publicly, each party performs a displacement on their mode.
  Finally, each mode is subject to squeezing.
  
  As both QND interaction and squeezing are the same on each mode,
  these operations commute with the $N$-port beam splitter for effective state preparation of Fig.~\ref{fig.scheme}.
  In terms of the separable modes of the state preparation scheme, the overall transformation reads
  (note that displacements do not have any effect on the variances)
  \begin{eqnarray}
    V_x &\to& a\left(V_x-\frac{g^2V_x^2}{g^2V_x+1}\right),\label{eq.qnd_Vx}\\
    V_p &\to& \frac{1}{a}(V_p+g^2),\label{eq.qnd_Vp}
  \end{eqnarray}
  and similarly for $W_x$, $W_p$.
  Here, $g$ is the strength of the QND interaction and $a$ gives the squeezing.
  Similarly as with noise addition strategy, requiring variance and correlation ratio $k'_1 = n'/m'$, $k'_2 = d'/c'$
  at the output, we get the set of equations
  \begin{eqnarray}
    \pi_x\delta_p g^4+\sigma_x\delta_p g^2+k'_2\delta_xa^2 &=& -\delta_p\\
    N\pi_x g^6+ (N\sigma_x+\pi_x\nu_p)g^4 +\nonumber\\+ (N+\sigma_x\nu_p)g^2 - Nk'_1\pi_xg^2 a^2 - k'_1\nu_xa^2 &=& -\nu_p,
  \end{eqnarray}
  where $\delta_x = V_x-W_x$, $\nu_x = (N-1)V_x+W_x$, $\pi_x = V_xW_x$, $\sigma_x = V_x+W_x$, and quantities with subscript $p$ are defined similarly.
  This set of equations can be solved analytically, e.g., by expressing $a^2$ from the first equation and plugging it into the second equation,
  leading to a cubic equation for $g^2$.

  Entanglement classification with QND interaction protocol for different values of variance and correlation ratio $k'_{1,2}$
  is shown in Fig.~\ref{fig.classes} (f)-(h).
  Compared to the correlated noise addition strategy, entanglement class is preserved for every input state.
  In addition, a larger subset of entangled states can be transformed for each set of protocol parameters
  than with the noise addition protocol.
  Finally, situation for different values of input variances $m$, $n$ is qualitatively the same.

  \section{Assisted quantum teleportation with symmetric Gaussian states}\label{sec.teleportation}

\begin{figure}
    \centering
    \includegraphics[width=\linewidth]{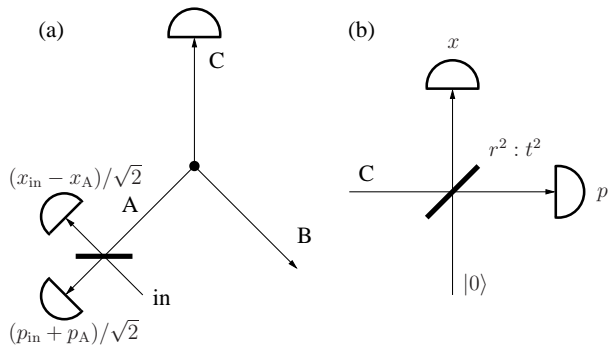}
    \caption{\label{fig.teleportation}
      (a) Scheme of assisted quantum teleportation with three parties.
      Alice performs a Bell measurement on mode A of the entangled state with an unknown coherent state in mode ``in'' she wants to teleport to Bob
      and announces the measurement result.
      Charlie helps to maximize the teleportation fidelity by performing certain measurement on mode C and announcing the result.
      (b) Detail of Charlie's measurement.
      Charlie splits mode C on a beam splitter with amplitude transmittance $t$ and reflectance $r$ and performs homodyne measurements of the outputs.
      In the limit $t\to 0$, he recovers homodyne measurement of the $x$ quadrature while for $t\to 1$, $p$ quadrature of the mode C is measured.
      Finally, $t = 1/\sqrt{2}$ corresponds to a heterodyne measurement of mode~C.
    }
  \end{figure}

  While entanglement classification is useful in order to understand the qualitative features of quantum states
  and the transformations we presented,
  for any practical applications of entangled states a figure of merit is required
  that would characterize how well given task can be performed with given state.
  Most generally, entanglement can be quantified by entanglement measures
  some of which have been proposed for multipartite Gaussian states~\cite{Adesso06,Adesso12};
  nevertheless, their calculation requires, in case of mixed states, taking Gaussian convex roof, 
  making the calculation rather cumbersome.
  Therefore, we use an operationalistic approach and are interested in fidelity of assisted quantum
  teleportation~\cite{Loock00,Yonezawa04}, as shown in Fig.~\ref{fig.teleportation}.
  Alice, who wants to teleport an unknown coherent state to Bob, performs a Bell measurement on the teleported state and her mode of the entangled state.
  Other parties help to improve the teleportation fidelity by certain measurements on their modes.
  Finally, all measurement results are publicly announced so that Bob can perform suitable displacements and get the teleported state.

  To find optimal measurement for the third party, Charlie, in case of quantum teleportation with three parties, we proceed as follows:
  We start from the covariance matrix of the tripartite state
  \begin{equation}
    \gamma_\mathrm{ABC} = \left(\begin{array}{cccccc} m&0&c&0&c&0 \\ 0&n&0&-d&0&-d \\ c&0&m&0&c&0 \\ 0&-d&0&n&0&-d \\
      c&0&c&0&m&0 \\ 0&-d&0&-d&0&n \end{array}\right)
  \end{equation}
  to which we add a fourth mode D in the vacuum state, $\gamma_\mathrm{ABC}\oplus I_\mathrm{D}$.
  Modes C and D are subsequently mixed on a beam splitter with transmittance $t$ and amplitude (phase) quadrature of mode C (D) is then measured.
  Thus, we can write the resulting covariance matrix of the modes A, B in block form
  \begin{eqnarray}
    \gamma_\mathrm{AB} &=& \left(\begin{array}{cc} A&C \\ C^T&B \end{array}\right), \\
    A = B &=& \diag\left(m-\frac{c^2r^2}{mr^2+t^2},n-\frac{d^2t^2}{nt^2+r^2}\right), \\
    C &=& \diag\left(c-\frac{c^2r^2}{mr^2+t^2},-d-\frac{d^2t^2}{nt^2+r^2}\right).
  \end{eqnarray}

  The teleportation fidelity can be expressed as~\cite{Fiurasek02}
  \begin{equation}
    F = \frac{2}{\sqrt{\det E}},
  \end{equation}
  where $E = 2D+RAR^T+RC+C^TR^T+B$, $D = I$ is the covariance matrix of the teleported coherent state,
  and, assuming $c>0$, $d>0$ (this corresponds to Alice measuring $x_\mathrm{in}-x_\mathrm{A}$, $p_\mathrm{in}+x_\mathrm{A}$), $R = \diag(-1,1)$.
  Using $t^2 = T$, $r^2 = 1-T$, we find that $\partial\det E/\partial T<0$ for $T\in[0,1]$;
  the optimal choice is $T = 1$, corresponding to a homodyne measurement of $p$ quadrature by Charlie.
  For the teleportation fidelity, we thus get
  \begin{equation}\label{eq.fidelity}
    F = \frac{1}{\sqrt{(m-c+1)(n-d+1-2d^2/n)}}.
  \end{equation}
  By comparison with teleportation in the bipartite case, $F = 1/\sqrt{(m-c+1)(n-d+1)}$, [by virtue of generalizing the result in Ref.~\cite{Fiurasek12}],
  we see that Charlie's measurement maximizes anti-correlations in Alice's and Bob's phase quadratures while preserving correlations in $x$;
  while measuring small portion of Charlie's amplitude quadrature would lead to increasing Alice's and Bob's $x$-correlations,
  it would also reduce the anti-correlations in their $p$ quadratures.

  \subsection{Correlated noise addition}
  
  Let us turn our attention to the scheme based on adding correlated noise and its effect on the teleportation fidelity.
  First important observation is that the amount of added noise does not influence the teleportation protocol
  as can be seen by noting that the noise variance enters the transformation formulas solely in Eq.~\eq{eq.noise_wx}.
  As a result, its contributions in the expressions for amplitude variance and amplitude correlations Eqs.~\eq{eq.m}, \eq{eq.c}
  cancel each other in the formula for fidelity Eq.~\eq{eq.fidelity}.
  In other words, while adding correlated noise increases the inter-modal amplitude correlations,
  it also leads to an increase of the amplitude variance, and these two effects exactly cancel each other.
  This holds also in the bipartite case; then, the fidelity takes the form $F = 1/\sqrt{(m'-c'+1)(n'-d'+1)}$, 
  keeping the crucial term $m'-c'$.
  This suggests that the finding is of a more general nature---while 
  it is necessary to keep both noise addition and squeezing to obtain an arbitrary combination of variance and correlation ratios
  [but, naturally, within bounds given by physicality of the expressions \eq{eq.noise_squeezing}, \eq{eq.noise_noise}],
  only the squeezing is responsible for the increase in teleportation fidelity.

  The optimal squeezing for a tripartite resource state can be found
  by taking the derivative of the argument of the square root in Eq.~\eq{eq.fidelity} $\partial[(m'-c'+1)(n'-d'+1-2d'^2/n')]/\partial a$ and putting it equal to zero.
  A straightforward calculation leads to the result
  \begin{equation}
    a_\mathrm{opt}^2 = \frac{V_x(2V_p+W_p)}{3V_pW_p}.
  \end{equation}
  In addition, it is easy to check that the second derivative is positive;
  hence the expression $(m'-c'+1)(n'-d'+1-2d'^2/n')$ reaches its minimum and the fidelity is maximal.

  In an experimental realization, however, it is not necessary to perform squeezing on all modes of the entangled state.
  Squeezing of Charlie's mode is directly followed by a homodyne measurement;
  hence, it can be emulated by properly rescaling the measurement result.
  Similarly, rescaling the result of Alice's Bell measurement corresponds to squeezing both of her modes.
  In addition, if both squeezing parameters are the same,
  the squeezing operations can be propagated through the balanced beam splitter, as discussed in Sec.~\ref{sec.math}.
  As a result, proper rescaling of Alice's measurement outcomes (without squeezing her mode of the entangled state) 
  corresponds to teleporting squeezed version of the input state.
  Therefore, instead of Alice and Charlie performing squeezing on their respective modes,
  Bob can equivalently rescale the classical signals corresponding to their measurement outcomes
  to obtain a squeezed version of the input state.
  Applying inverse squeezing (either on the mode or on his measurement data),
  Bob can recover the original input coherent state.

\begin{figure}
    \centering
    \includegraphics[width=\linewidth]{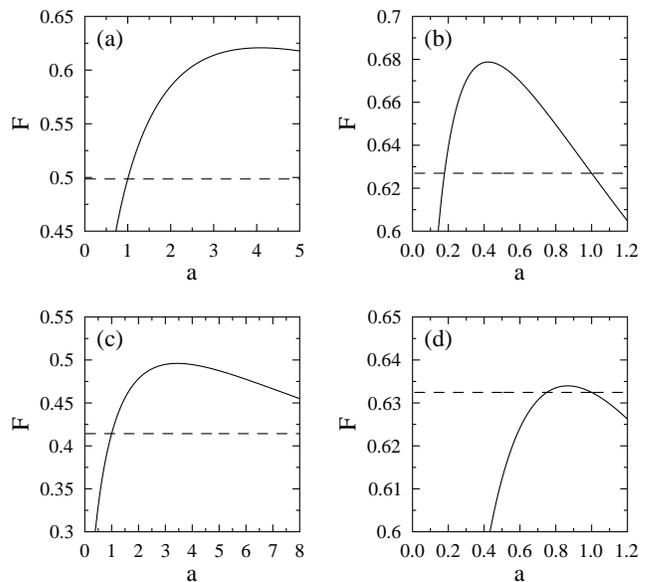}
    \caption{\label{fig.squeezing}
      Teleportation fidelity $F$ as a function of squeezing $a$ for states with $(m,n,c,d) = (7.5,7.5,5,3.7)$ (a),
      $(4,4,3.8,1.6)$ (b), $(4,4,0.5,1.9)$ (c), $(4,1.5,3.5,0.5)$ (d).
      The dashed lines show the fidelity of the original state, the full lines give fidelity after squeezing.
      Best fidelity improvement can be reached with states containing more initial noise (a).
      While, in this case, the fidelity is initially below the classical limit of 0.5,
      local squeezing can improve the fidelity by about 25~\%, leading to a value of 0.62.
    }
  \end{figure}

  To demonstrate the power of squeezing to improve teleportation fidelity, in Fig.~\ref{fig.squeezing},
  we study the teleportation fidelity as a function of squeezing for several resource states.
  Comparison with initial fidelity (dashed line in Fig.~\ref{fig.squeezing})
  shows that the best improvement of teleportation fidelity can be achieved with states with high initial noise [Fig.~\ref{fig.squeezing} (a)];
  this conjecture has been supported by extensive numerical calculations.
  The optimal squeezing is, in the case of Fig.~\ref{fig.squeezing} (a), about 6~dB which would be very difficult to realize perfectly in an experiment;
  note that general noise introduced by imperfect squeezing would, in contrast to correlated noise added to amplitude quadratures,
  lead to a drop in the teleportation fidelity.
  Nevertheless, the curve in Fig~\ref{fig.squeezing} (a) is rather flat around the maximum
  making it possible to use lower squeezing without much decrease in fidelity.
  Even with 3 dB of squeezing, it is possible to reach an enhancement of almost 20~\%.
  In general, numerical calculations suggest that the higher the optimum squeezing is,
  the wider the maximum is, making it possible to use smaller values of squeezing and still achieve significant increase in teleportation fidelity.

  \subsection{QND interaction}

  A general analysis is more complicated in case of the protocol with partial QND measurement.
  In this case, it is not possible to find a closed formula for optimal interaction strength $g$.
  Nevertheless, if we use the transformation formulas \eq{eq.qnd_Vx}-\eq{eq.qnd_Vp}
  and plug them into formulas for the covariance matrix parameters \eq{eq.m}-\eq{eq.d}
  it can be shown that $F\to 0.5$ for $g\to\infty$ independent of the resource state.
  This follows by finding QND-interaction-strength-dependent optimum squeezing $a_\mathrm{opt}$ from
  $\partial[(m'-c'+1)(n'-d'+1-2d'^2/n')]/\partial a = 0$
  and taking the limit to arrive at $\lim_{g\to\infty}(m'-c'+1)(n'-d'+1-2d'^2/n') = 4$.
  Similarly, we can reach a local extremum in fidelity by setting $g = 0$.
  Extensive numerical calculations suggest that these are the only two possible maxima of the teleportation fidelity as a function of the interaction strength.

\begin{figure}
     \centering
     \includegraphics[width=\linewidth]{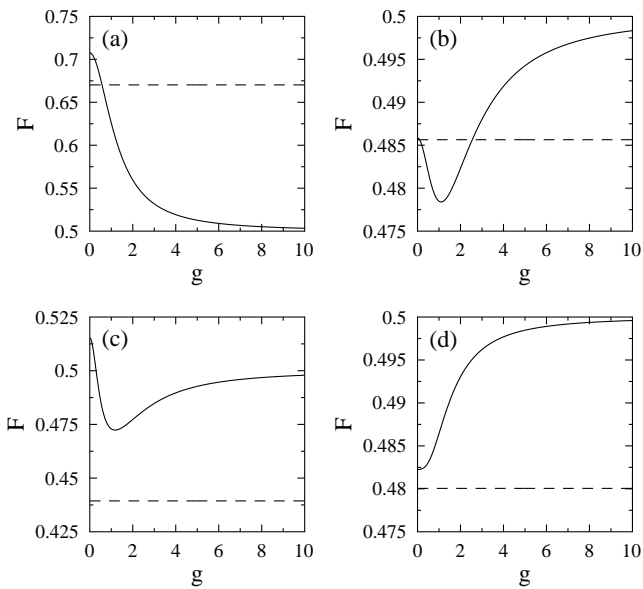}
     \caption{ \label{fig.qnd_fidelity}
        Dependence of the teleportation fidelity $F$ on the QND interaction strength $g$ for states with
	$(m,n,c,d) = (4,4,3.8,1.7)$ (a), $(4,4,3,1.6)$ (b), $(4,4,1,1.9)$ (c), $(1.5,1.5,0.2,0.4)$ (d).
	The dashed lines show the fidelity with the original resource states.
	Generally, local maxima occur only for $g = 0$ or $g\to\infty$.
     }
  \end{figure}

  To illustrate this point, in Fig.~\ref{fig.qnd_fidelity}, we plot the teleportation fidelity as a function of interaction strength for several resource states.
  While there can, in general, be a local minimum for a finite nonzero interaction strength [such as in Fig.~\ref{fig.qnd_fidelity} (b), (c)],
  only $g = 0$ or $g\to\infty$ can play the role of a local maximum.
  If the former choice is true, this corresponds to improving the fidelity by squeezing alone, as discussed in the previous section;
  the latter option leads to a projective measurement of the multipartite resource state, followed by a measure-and-prepare strategy
  which cannot exceed the $F = 0.5$ bound and is, for this reason, of little interest.

  \section{Conclusions}\label{sec.conclusions}

  In summary, we investigated two protocols for transformations of permutation symmetric Gaussian states
  by means of local Gaussian operations and classical communication.
  In the transformations, we were interested in keeping permutation invariance of the state
  while changing the ratio of amplitude and phase variances and correlations.

  While the transformation parameters are determined generally for any number of parties,
  our numerical analysis is focused on tripartite states
  as they constitute the simplest class of multipartite states in terms of entanglement classification,
  having three entanglement classes permutation symmetric Gaussian states can fall into---fully entangled, bound entangled
  or fully separable states.

  Firstly, we considered a protocol based on adding correlated noise followed by local squeezing operation.
  In this setup, states with correlations in amplitude and anti-correlations in phase quadratures
  keep their entanglement class, apart from a very small subset of fully entangled states that become bound entangled,
  if the noise is added in the amplitude quadratures.
  This does not hold for states with anti-correlations in amplitude,
  since adding correlated noise to the anti-correlated quadrature decreases the anti-correlations.
  As a result, the entanglement is degraded for states with correlations in phase and anti-correlations in amplitude;
  nevertheless, this can be avoided if one adds the noise to the phase quadratures instead.

  Next, we used quantum non-demolition interaction with ancillary vacuum modes and local squeezing.
  This approach is experimentally more challenging
  (QND interaction can be achieved by interaction of light modes with atomic ensembles~\cite{Hammerer10}
  or in cavity quantum optomechanics setups~\cite{Aspelmayer13},
  or it can be emulated using linear optics and additional modes~\cite{Filip05})
  but this approach has two major advantages compared to the noise addition scheme.
  First of all, this quantum filtering protocol preserves entanglement classes for all input states
  and second, for given values of variance and correlation ratio, a larger subset of fully entangled states can usually be
  transformed, indicating a better applicability of this protocol.
  Similar to the noise addition strategy, the possibilities of the protocol can be improved by considering
  partial QND measurement of the phase quadrature.

  Apart from the transformation protocols, we also introduced an experimental scheme
  for generation of permutation symmetric Gaussian states.
  This setup is relevant not only from experimental point of view
  but also because it significantly simplifies description of studied states, and theoretical analysis of the protocols.
  In fact, this approach can in future be used to assess properties of other feasible protocols
  for manipulations of permutation invariant Gaussian states.

  Finally, we also investigated the fidelity of assisted quantum teleportation with permutation symmetric Gaussian states
  to get a quantitative characterization of the change of the structure of the multipartite entanglement.
  For correlated noise addition strategy, we showed that the amount of added noise does not affect the fidelity at all
  and optimum squeezing that maximizes the teleportation fidelity can be found.
  On the other hand, in the QND interaction protocol, the fidelity depends on the interaction strength.
  We identified two possible fidelity maxima---for zero interaction strength,
  meaning there is no non-demolition interaction with an ancillary system,
  and for infinitely strong interaction,
  corresponding to a projective measurement and a classical measure-and-prepare teleportation strategy;
  our numerical results suggest that there are no other possible maxima for teleportation with tripartite states.
  Thus, while the QND interaction protocol serves better than correlated noise addition strategy
  when only entanglement classification is concerned, it degrades the teleportation fidelity.
  While teleportation fidelity cannot be viewed as an entanglement measure
  as it can be deterministically affected by local operations and classical communication,
  it would be interesting to investigate whether similar result holds also for entanglement measures
  (such as R\'enyi entropy of order 2~\cite{Adesso12} or the contangle~\cite{Adesso06,Adesso06b})
  or for other applications of continuous-variable entanglement.

  \begin{acknowledgments}
    This work was supported by the Czech Science Foundation (Grant No. P205/12/0694) and by the Palack\'y University (Project No. PrF-2013-008).
  \end{acknowledgments}

  \bibliography{fidelity}

\end{document}